\documentclass[reprint,amsmath,amssymb,aps]{revtex4-1}
\renewcommand{\v}[1]{\ensuremath{\mathbf{#1}}} 
\newcommand{\gv}[1]{\ensuremath{\mbox{\boldmath$ #1 $}}} 
\newcommand{\avg}[1]{\left< #1 \right>} 
\renewcommand{\d}[2]{\frac{d #1}{d #2}} 
\newcommand{\pd}[2]{\frac{\partial #1}{\partial #2}} 
 
\newcommand{\grad}[1]{\gv{\nabla} #1} 
\let\baraccent=\= 
\renewcommand{\=}[1]{\stackrel{#1}{=}} 

\newcommand{\p}{^{\prime} }
\newcommand{\comment}[1]{}

\usepackage{graphicx}
\usepackage{dcolumn}
\usepackage{multirow}
\usepackage{natbib}

\begin{document}
\def\reff@jnl#1{{\rm#1}} 

\def\araa{\reff@jnl{ARA\&A}}             
\def\aj{\reff@jnl{AJ}}                   
\def\apj{\reff@jnl{ApJ}}                 
\def\apjl{\reff@jnl{ApJ}}                
\def\apjs{\reff@jnl{ApJS}}               
\def\apss{\reff@jnl{Ap\&SS}}             
\def\aap{\reff@jnl{A\&A}}                
\def\aapr{\reff@jnl{A\&A~Rev.}}          
\def\aaps{\reff@jnl{A\&AS}}              
\def\baas{\ref@jnl{BAAS}}               
\def\jcap{\reff@jnl{J. Cosmology Astropart. Phys.}}
\def\jrasc{\reff@jnl{JRASC}}             
\def\memras{\reff@jnl{MmRAS}}            
\def\mnras{\reff@jnl{MNRAS}}             
\def\na{\reff@jnl{New A}}                
\def\nar{\reff@jnl{New A Rev.}}          
\def\pra{\reff@jnl{Phys.~Rev.~A}}        
\def\prb{\reff@jnl{Phys.~Rev.~B}}        
\def\prc{\reff@jnl{Phys.~Rev.~C}}        
\def\prd{\reff@jnl{Phys.~Rev.~D}}        
\def\pre{\reff@jnl{Phys.~Rev.~E}}        
\def\prl{\reff@jnl{Phys.~Rev.~Lett.}}    
\def\pasa{\reff@jnl{PASA}}               
\def\pasp{\reff@jnl{PASP}}               
\def\pasj{\reff@jnl{PASJ}}               
\def\qjras{\reff@jnl{QJRAS}}             
\def\physrep{\reff@jnl{Phys.~Rep.}}   
\let\astap=\aap
\let\apjlett=\apjl
\let\apjsupp=\apjs
\let\applopt=\ao
\preprint{APS/123-QED}

\title{A Multi-Wavelength Analysis of Annihilating Dark Matter as\\ the Origin of the Gamma-Ray Emission from M31}
\author{Alex McDaniel}
 \email{alexmcdaniel@ucsc.edu}
\author{Tesla Jeltema}
 \email{tesla@ucsc.edu}
 \author{Stefano Profumo}
 \email{profumo@ucsc.edu}
\affiliation{
Department of Physics, University of California,
1156 High Street, Santa Cruz, California, 95064, USA
}
\affiliation{
Santa Cruz Institute for Particle Physics,
1156 High Street, Santa Cruz, California, 95064, USA
}
\date{\today}
\begin{abstract}
Indirect detection of dark matter (DM) by multi-wavelength astronomical observations provides a promising avenue for probing the particle nature of DM. In the case of DM consisting of Weakly-Interacting Massive Particles (WIMPs), self-annihilation ultimately produces various observable products including $e^{\pm}$ pairs and gamma-rays. The gamma-rays can be detected directly, while the $e^{\pm}$ pairs can be detected by radio emission from synchrotron radiation or X-rays and soft gamma-rays from inverse Compton scattering. An intriguing region to search for astrophysical signs of DM is the Galactic center (GC) of the Milky Way, due in part to an observed excess of gamma-rays that could be DM. A recent observation by the Fermi-LAT collaboration of a similar excess in the central region of the Andromeda galaxy (M31) leads us to explore the possibility of a DM-induced signal there as well. We use the RX-DMFIT tool to perform a multi-frequency analysis of potential DM annihilation emissions in M31. We consider WIMP particle models consistent with the GC excess and calculate the expected emission across the electromagnetic spectrum in comparison with available observational data from M31. We find that the particle models that best fit the M31 excess favor lower masses than the GC excess. The best fitting models are for a $b\bar{b}$ final state with $M_{\chi}=11$ GeV and $\avg{\sigma v}=2.6\times 10^{-26}$ cm$^3$s$^{-1}$, as well as an evenly mixed $b\bar{b}/\tau^+\tau^-$ final state with $M_{\chi}=5.8$ GeV and $\avg{\sigma v}=2.03\times 10^{-26}$ cm$^3$s$^{-1}$. For conservative estimates of the diffusion and magnetic field models the expected radio emissions appear to be in tension with currently available data in the central region of M31, although this constraint has a fairly strong dependence on the values chosen for parameters describing the magnetic field strength and geometry.
\end{abstract}

\pacs{Valid PACS appear here}
\maketitle

\section{\label{sec:intro}Introduction}
Understanding the fundamental nature of dark matter remains one of the foremost problems in physics. Particle dark matter is arguably the best-supported explanation, and Weakly-Interacting Massive Particles (WIMPs) in particular have strong theoretical motivation as a potential candidate \cite{bertone, jungman, bergstrom}. One of several possible methods for testing WIMP models of dark matter is through indirect detection using astrophysical observations. In the case of annihilating dark matter, the byproducts of dark matter annihilation can include standard model particles such as quarks, leptons, and bosons, which then decay into particles that can be detected through a variety of observational experiments. For instance, the production of electrons and positrons can produce radio emission through synchrotron radiation in regions where magnetic fields are present, or X-rays and soft gamma-rays by up-scattering ambient photons through inverse Compton scattering. Additionally, dark matter annihilation is expected to produce prompt gamma-rays predominately from neutral pion ($\pi^0$) decay that have been a major focus of dark matter indirect detection searches. While the bulk of indirect searches for annihilating dark matter have been performed by studying these gamma-rays, several studies have shown that the radio \cite{storm, storm16, M31USC, chan, laha, RegisUllio, Natarajan, spekkens,  natarajanSegue, colaDraco} as well as x-ray \cite{JP2012, TeslaXray, RegisUllio} approaches have the potential to place competitive and in some cases stronger constraints on dark matter in a variety of systems including galaxy clusters, dwarf galaxies, and the central regions of normal galaxies.

An especially enticing target for indirect dark matter searches is our own Galactic center, which has been widely studied in the context of dark matter due in part to its proximity as well as its high concentration of dark matter. Additionally, the presence of a gamma-ray excess in the inner galaxy known as the Galactic Center Excess (GCE) has been reported by several groups using Fermi-LAT data and could potentially be explained as a dark matter signal from annihilating WIMPs \cite{goodenough, abazajian2014, calore, daylan, GordonMacias, HooperGoodenough} (or for a review see \cite{FermiReview}). Other possible explanations for the GCE that have been explored include an unresolved population of millisecond pulsars (MSPs) \cite{EcknerMSP, bartels, brandt} or additional cosmic ray sources \cite{carlson, gaggero, cholis}. Several of the dark matter interpretations have been shown to be consistent with observations for certain WIMP models, specifically for those annihilating through $b\bar{b}$ and $\tau^+\tau^-$ channels with masses of $\sim 30-50$ GeV and $\sim 7-10$ GeV, respectively (see e.g. \cite{abazajian2014, calore, daylan, GordonMacias, HooperGoodenough}).

A similar excess in the nearby Andromeda galaxy (M31) has been reported by the Fermi collaboration \cite{FermiM31}. The dataset used in the analysis by Ackermann et. al. (2017) \cite{FermiM31} includes 88 months of Pass 8 data collected between August 4, 2008, and December 1, 2015. SOURCE class events were used excluding those with zenith angle greater than $90^{\circ}$ or rocking angle greater than $52^{\circ}$. Reconstructed events within an energy range 0f 0.1-100 GeV were considered as well as reconstructed directions within a $14^{\circ}\times14^{\circ}$ region of interest (ROI) centered at $(\alpha,\delta) =  (10^{\circ}.6847, 41^{\circ}.2687 )$. For greater detail about the analysis we refer the reader to Ref.~\cite{FermiM31}.

The results from this analysis motivate an examination of similar possible explanations of the observed excess as in the case of the GCE. Already there has been exploration into the MSP explanation for the M31 excess \cite{EcknerMSP}, where the point was made that MSPs are unlikely to be able to account for the entirety of the observed emission. In this paper, we use the recently developed RX-DMFIT \cite{McDaniel} tool to explore dark matter annihilation as a potential source of the observed excess, and consider the multi-wavelength emissions that would be expected due to synchrotron radiation, inverse Compton scattering of CMB and starlight photons, and $\pi^0$ decay and other prompt gamma-rays from dark matter annihilation. We particularly focus on the radio and gamma-ray aspects as they provide greater insight than the X-rays into the dark matter interpretation with current observational data. M31 has been the focus of some previous radio-only dark matter studies \cite{M31USC, chan} as well as an analysis comparing the DM induced gamma-ray emission in M31 with multi-wavelength emission in other systems \cite{colaM31}. Here however we study the full spectrum expected from dark matter annihilation in M31 and we compare directly with data available in the literature in order to provide complementary probes of a dark matter interpretation for the gamma-ray emission from Andromeda. Our analysis thus provides a two pronged approach wherein we seek to determine whether the GCE dark matter particle models provide gamma-ray emissions consistent with the M31 observations, as well as whether potential dark matter particle models that could explain the M31 gamma-ray excess also predict radio and X-ray emissions ``self-consistent'' with available M31 observations. 

This paper is organized in the following manner. In section \ref{sec:astro} we present our astrophysical model for M31, including relevant astrophysical model components such as the diffusion model parameter, the magnetic field, the dark matter density profile, and the interstellar radiation field (ISRF). We derive expressions for the synchrotron and inverse Compton emissions from DM annihilation in section \ref{sec:emission}, then describe our particle physics models in section \ref{sec:particle}. Our results comparing the expected emissions due to dark matter annihilation and the observational data are presented in section \ref{sec:results}, and finally we conclude in section \ref{sec:conclusion}. Throughout this paper we assume a $\Lambda$CDM universe with $H_0 = 70.4$ km s$^{-1}$ Mpc$^{-1}$ , $\Omega_m = 0.27$, $\Omega_{\Lambda} = 0.73$.

\section{\label{sec:astro} Astrophysical Modeling}
\subsection{\label{sec:diffusion}Diffusion}
The computation of expected emissions due to the injection of electrons and positrons from DM annihilation requires solving a diffusion equation of the type
\begin{equation}\label{eq:diffusion}
\begin{split}
\pd{}{t} \pd{n_e}{E} = \grad \left [D(E,\v{r}) \grad{ \pd{n_e}{E} } \right] +
\pd{}{E}\left[b(E,\v{r}) \pd{n_e}{E} \right]\\
+Q(E,\v{r}),
\end{split}
\end{equation}
where $\partial n_e /\partial E$ is the electron equilibrium spectrum and the source term from DM annihilation, $Q(E,\v{r})$, is given by 
\begin{equation}\label{eq:source}
Q\left(E, r\right) = \frac{\avg{\sigma v } \rho_{\chi}^2(r) }{2M_{\chi}^2 }\sum_f BR_{f}\d{N}{E_{inj}},
\end{equation}
where $\rho_{\chi}(r)$ is the DM density profile, $M_{\chi}$ is the DM mass, $\avg{\sigma v}$ is the thermally averaged annihilation cross-section, and $dN/dE_{inj}$ is the $e^{\pm}$ injection spectrum through annihilation channels with branching ratios $BR_{f}$. The equation above makes several simplifying assumptions, including the absence of diffusive reacceleration and convection, and is well-defined once boundary conditions are specified; also, the factor 2 in the denominator of Eq.~(\ref{eq:source}) implicitly assumes that the dark matter is its own antiparticle. 

In the energy loss term $b(E,\v{r})$ we include contributions from synchrotron radiation, inverse Compton scattering of CMB and starlight photons, Coulomb interactions, and bremsstrahlung radiation, given by the expression
\begin{equation}\label{eq:bloss}
\begin{split}
b(E,\v{r}) & = b_{IC}(E) + b_{Synch.}(E,\v{r}) + b_{Coul.}(E) + b_{Brem.}(E)\\
	& = b_{IC}^0u_{CMB}E^2 + b_{IC}^0u_{SL}E^2 + b^0_{Synch.}B^2(r) E^2 \\
	&+ b_{Coul.}^0 n_e \left(1+\log\left(\frac{E/m_e}{n_e }\right)/75 \right)\\
	& + b^0_{Brem.}n_e  \left( \log\left(\frac{E/m_e}{n_e }\right) + 0.36 \right).
\end{split}
\end{equation}
where the constants are in units of $10^{-16}$ GeV s$^{-1}$ and have values $b_{syn}^0 \simeq 0.0254$, $b_{IC}^0 \simeq 0.76$, $b_{brem}^0 \simeq 1.51$, and $b_{Coul}^0 \simeq 6.13$ \cite{longair, cola}. Additionally, we take the photon energy densities to be $u_{SL} = 8$ eV cm$^{-3}$ for starlight and $u_{CMB} = 0.25$ eV cm$^{-3}$ for CMB photons \cite{ProfumoUllio, porter}. We will work under the assumption of a steady-state solution, and we thus set the left hand side to zero while noting that the analytic solution can be determined in the case on non-stationary sources \cite{cola, ginzburg}; also, we adopt a homogeneous diffusion coefficient of the form,
\begin{equation}\label{eq:Dcoeff}
D(E) = D_0 E^{\delta}.
\end{equation}
The similarity between the Andromeda galaxy and the Milky Way motivates us to adopt Galactic diffusion parameter values; we thus employ $D_0 =3\times 10^{28}$ cm$^{2}$s$^{-1}$ and $\delta = 0.3$, which are representative values for the Milky Way \cite{SMP, vladimirov, BaltzEdsjo, WLG}. While previous analyses of DM annihilation in Andromeda neglect diffusion \cite{M31USC, chan}, we take this into account for a more conservative analysis, noting in particular that diffusion is relevant on the smaller scales that we explore in this work. There are also other potential astrophysical processes that could depress the signal such as convection and reacceleration \cite{SMP, winds}, however we do not consider these effects in this paper. The full analytic solution to the diffusion equation with free-escape boundary conditions is \cite{cola, ginzburg},
\begin{multline}
G(r,  \Delta v ) = \frac{1}{\sqrt{4\pi \Delta v }}\sum_{n = -\infty}^{\infty} \left(-1\right)^n\int_0^{r_h}dr\p\frac{r\p}{r_n} \left(\frac{ \rho_{\chi}(r\p) }{\rho_{\chi}(r)}\right)^2\\ 
 \times\left[
\exp\left(-\frac{ (r\p - r_n)^2 }{4 \Delta v}	\right)	- 	\exp\left(-\frac{ (r\p + r_n)^2 }{4 \Delta v}	\right)	\right],
\end{multline}
where $r_h$ is the diffusion zone radius and the locations of the image charges used to implement the free-escape boundary condition are $r_n = \left(-1\right)^n + 2nr_h$. The value $\Delta v$ is defined as $\Delta v = v(E) - v(E\p)$ with 
\begin{equation}
v(E) = \int_E^{M_{\chi}} d\tilde{E} \frac{D(\tilde{E})}{b(\tilde{E})}.
\end{equation}
where we have adopted the spatially independent form of the energy loss term as previously described \cite{McDaniel}.

\subsection{\label{sec:Bfield} Magnetic Field}
The predicted synchrotron emission from the $e^{\pm}$ products of dark matter annihilation depends heavily on the magnetic field strength and profile. While the full three dimensional structure of the magnetic fields in the central region of M31 can be highly complex, the magnitude of the fields as determined by Faraday rotation measures of polarized radio emission have been reported to have strengths of $15 \pm 3 \: \mu$G for $r = 0.2-0.4$ kpc and $19 \pm 3 \: \mu$G for $r = 0.8-1$ kpc \cite{giessubel, hoernes}, whereas the regular magnetic field in the outer regions is fairly constant with a typical strength of roughly $5 \pm \: 1 \mu$G \cite{fletcher}. In this study we model the magnetic field of M31 with an exponential component as well as a constant component with the form,
\begin{equation}\label{eq:bfield}
B(r) = B_0e^{-r/r_c} +  B_{const}.
\end{equation}
For consistency with above quoted values, we adopt $B_0 = 10 \: \mu$G and $B_{const} = 5 \: \mu$G, as well as taking $r_c = 1.5$ kpc based on estimates of the magnetic field scale height \cite{fletcher}. Since we are only interested in a relatively small region of radius $r  \sim 1 - 5$ kpc, we assume a spherical magnetic field model while acknowledging that a more accurate model of the magnetic field would include another spatial dependence perpendicular to the plane of the galaxy in order to better model the disk structure at larger radii.

\subsection{\label{sec:dmProfile} Dark Matter Density Profile}
Previous studies have shown that the M31 rotation curves can be fit with good results by using a Navarro-Frenck-White (NFW) \cite{NFW96, NFW97} profile \cite{tamm, sofue}. Here we adopt a generalized NFW profile of the form
\begin{equation}\label{dmprofile}
\rho_{\chi}(r) = \frac{\rho_s}{\left(\frac{r}{r_s}\right)^{\gamma}\left[1+\left(\frac{r}{r_s}\right)\right]^{3-\gamma}},
\end{equation}
where $\gamma$ is a free parameter. For a standard NFW profile we take $\gamma = 1$, however when including significant baryonic matter, such as in the central regions of galaxies, the DM distribution is expected to have a more centrally peaked profile \cite{BlumenthalFaber, RydenGunn}. Values used in GCE analysis are typically about $\gamma \sim 1.2 - 1.3$ \cite{calore, daylan, GordonMacias, HooperGoodenough}. Thus, we will examine a variety of $\gamma$ values, taking a default of $\gamma=1.25$. The values for the characteristic density, $\rho_s$, and scale radius, $r_s$, are taken to be $0.418$ GeV cm$^{-3}$ and 16.5 kpc respectively \cite{tamm}.

\subsection{\label{sec:ISRF} Interstellar Radiation Field}
In modeling the ISRF for Andromeda we include two elements: the CMB radiation field which is modeled exactly by a black-body spectrum with temperature $T=2.735$ K, as well as a starlight (SL) radiation component. We approximate the SL energy spectrum as a black-body with $T=3500$ K, following previous work showing this is a reasonable assumption for the case of the Milky Way \cite{CirelliIC}. Unlike the CMB radiation field that is constant throughout the universe, our SL radiation field requires including a spatial dependence. For this, we use a two component bulge-disk model that follows the luminosity profile of M31 \cite{courteau}. Specifically, for the bulge component we employ
\begin{equation}
n_b(r) \propto e^{-b_n\left[\left(\frac{r}{r_b}\right)^{1/n}-1\right]},
\end{equation}
and for the disk we employ instead
\begin{equation}
n_d(r) \propto e^{-\frac{r}{r_d}}
\end{equation}
The values for the various parameter $b_n,\ r_b,\ n,\ r_d$ 
are taken from Ref.~\cite{courteau}. From figure 9 of \cite{courteau} we estimate the ratio of the bulge luminosity to the disk luminosity in the central regions of M31 to be $\sim 1/135$. Thus, the spatial profile of our SL model is given by, 
\begin{equation}
n(r) \propto e^{-b_n\left[\left(\frac{r}{r_b}\right)^{1/n}-1\right]}+\frac{e^{-\frac{r}{r_d}}}{135}.
\end{equation}
Including both the spatial and spectral components we have 
\begin{equation}
n(\nu, r) = N\frac{8\pi \nu^2/c^3}{e^{h\nu/kT}-1}\left[ e^{-b_n\left[\left(\frac{r}{r_b}\right)^{1/n}-1\right]}+\frac{e^{-\frac{r}{r_d}}}{135} \right]
\end{equation}
where $N$ is a normalization factor. This factor is determined by assuming that the SL energy density in the central regions of M31 is similar to that of the MW, which is roughly $\sim 8$ eV cm$^{-3}$ \cite{porter, ProfumoUllio}, giving us a value of $N = 4.9\times 10^{-11}$. In table \ref{tab:Params} we summarize our parameter selection. These are the values used throughout our analysis unless otherwise noted. 

\begin{table}[tbph]
\centering
\def\arraystretch{1.5}
\begin{tabular}{cc}
\hline
\multicolumn{2}{c}{Astrophysical Parameters}\\
\hline \hline
$d$              &\qquad 780 kpc\\ 						
$r_h$          &\qquad 5 kpc\\						
$r_{ROI}$   &\qquad 5 kpc\\			
$r_{core}$ &\qquad 1.5 kpc\\ 		
$B(r)$         & \qquad$B_0e^{-r/rc} + B_{const}$\\
$B(0)$        & \qquad15 $\mu$G\\ 			
$\rho_{s}$ &\qquad 0.418 GeV/cm$^3$\\	
$r_{s}$        & \qquad16.5 kpc\\
$\gamma$ &\qquad 1.25\\
$D(E)$ & \qquad$D_0E^{\delta}$\\
$D_0$ & \qquad$3\times 10^{28}$cm$^2$ s$^{-1}$\\	
$\delta$ &\qquad 0.3\\ 
\hline \hline
\end{tabular}
\caption{\label{tab:Params}Default astrophysical parameters. These values are used throughout unless otherwise noted.}
\end{table}
\section{\label{sec:emission}Emission from Dark Matter Annihilation}
\subsection{\label{sec:synch}Synchrotron}
In addition to gamma-rays from prompt emission in the annihilation event, the injection of charged electrons and positrons from DM annihilation is expected to produce multi-wavelength emission through processes including synchrotron radiation, inverse Compton scattering of ambient photons, bremsstrahlung, and Coulomb interactions. In the presence of magnetic fields stronger than the equivalent CMB energy density ($B_{CMB} =3.25(1+z)^2\:\mu$G), synchrotron radiation is the dominant energy loss process of the electron/positron byproducts of DM annihilation. The synchrotron power for a frequency $\nu$ averaged over all directions is \cite{storm16, longair} 

\begin{equation}
P_{syn} \left(\nu, E , r\right) = \int_0^{\pi} d\theta \frac{\sin \theta}{2} 2\pi \sqrt{3}r_0 m_e c \nu_0 \sin\theta F\left(\frac{x}{\sin\theta}\right), 
\end{equation}
where $r_0 = e^2/(m_ec^2)$ is the classical electron radius, $\theta$ is the pitch angle, and $\nu_0 = eB/(2\pi m_e c)$ is the non-relativistic gyrofrequency. The $x$ and $F$ terms are defined as,

\begin{equation}
x \equiv \frac{2\nu \left(1+z\right)m_e^2}{3\nu_0 E^2},
\end{equation}

\begin{equation}
F(s) \equiv s\int_s^{\infty} d \zeta K_{5/3}\left( \zeta \right)1.25 s^{1/3}e^{-s}\left[648 + s^2\right]^{1/12},
\end{equation}
where $K_{5/3}$ is the Bessel function of order 5/3. The synchrotron emissivity is then

\begin{equation}
j_{syn} \left(\nu , r\right)= 2\int_{m_e}^{M_{\chi}} dE \d{n_{e}}{E}\left(E, r\right)P_{syn}\left(\nu, E, r \right).
\end{equation}
The integrated flux density spectrum can then be taken to be \cite{McDaniel, cola}
\begin{equation}\label{eq:ssyn}
S_{syn} \approx \frac{1}{D_A^2} \int dr r^2 j_{syn}(\nu, r ),
\end{equation}
where $D_A$ is the angular diameter distance.
\subsection{\label{sec:IC}Inverse Compton}
In addition to synchrotron radiation, inverse Compton scattering of ambient CMB and starlight photons is a significant radiative loss process for $> 1$ GeV electrons and positrons. Depending on the mass of the DM particle \cite{JP2012}, the upscattered CMB photons will peak in the soft to hard X-rays, and the higher energy SL photons will upscatter into the hard X-ray up to soft gamma-ray regime, with higher DM masses leading to higher energy resulting spectra in each case. 
With the photon number density $n(\epsilon, r) = n_{CMB}(\epsilon) + n_{SL}(\epsilon, r)$, and the IC scattering cross-section $\sigma\left( E_{\gamma} , \epsilon , E \right)$, the IC power is

\begin{equation}
P_{IC} \left( E_{\gamma}, E , r\right) = c E_{\gamma}\int d\epsilon \: n \left( \epsilon, r \right) \sigma\left( E_{\gamma} , \epsilon , E \right)
\end{equation}

\noindent
where $\epsilon$ is the energy of the target photons, $E$ is the energy of the relativistic electrons and positrons, and $E_{\gamma}$ is the energy of the photons after scattering. The scattering cross-section, $\sigma\left( E_{\gamma} , \epsilon , E \right)$, is given by the Klein-Nishina formula:
\begin{equation}
\sigma\left( E_{\gamma} , \epsilon , E \right) = \frac{3 \sigma_T}{4\epsilon \gamma^2} G\left( q, \Gamma \right),
\end{equation}

\noindent where $\sigma_T$ is the Thomson cross-section and $G(q, \Gamma)$ is given by \cite{Blumenthal}:

\begin{equation}
G (q, \Gamma) = \left[  2q\ln q + (1+2q)(1-q) + \frac{(2q)^2(1-q)}{2(1+\Gamma q)}  \right],
\end{equation}
where,
\begin{equation}
\Gamma = \frac{4\epsilon \gamma}{m_e c^2} = \frac{4\gamma^2 \epsilon}{E}, \qquad q = \frac{E_{\gamma}}{\Gamma(E-E_{\gamma})} 
\end{equation}
The kinematics of inverse Compton scattering set the range of $q$ to be $1/\left( 4 \gamma^2 \right) \leq q \leq1$ \cite{cola, Blumenthal, Rybicki}.
As with the synchrotron flux calculation, the local emissivity is
\begin{equation}
j_{IC} \left(E_{\gamma}, r\right)= 2\int_{m_e}^{M_{\chi}} dE \d{n_{e}}{E}\left(E, r\right)P_{IC}\left(E, E_{\gamma}\right),
\end{equation}
and the integrated flux density is:
\begin{equation}
 S_{IC} \approx \frac{1}{D_A^2} \int dr r^2 j_{IC}(E_{\gamma}, r ).
\end{equation}

\subsection{Gamma-rays}
Calculating the gamma-ray flux from DM annihilation is straightforward in comparison to the synchrotron and IC fluxes since gamma-ray photons do not undergo the same radiative loss and diffusion processes. For the gamma-ray flux we integrate over the source volume \cite{cola, RegisUllio}, 
\begin{equation}
F_{\gamma}= \frac{1}{D_A^2}\int dr  r^2 E^2Q_{\gamma}\left(E, r\right).
\end{equation}

\section{\label{sec:particle}Particle Physics Framework}
In this analysis we seek to (i) test whether the hypothesis that the gamma-ray emission from M31 originates from dark matter annihilation is compatible with the same explanation to the Galactic center excess (GCE) and (ii) examine whether WIMP dark matter can explain the gamma-ray excess observed in M31 by the Fermi collaboration \cite{FermiM31} compatibly with observations of M31 at other wavelengths. 
While task (ii) doesn't necessarily entail specific choices for the pair-annihilation final state of the dark matter, given the relatively meager spectral information on the gamma-ray emission from M31, it does provide us with a preferred range for the particle dark matter mass. In order to pursue task (i) and (ii) simultaneously, and for simplicity, we shall consider annihilation final states that have been suggested as plausible candidates to describe the GCE from the standpoint of the reconstructed GCE spectrum. 

Specifically, following the results of previous studies of the GCE we focus on particle models with $M_{\chi} = 40$ GeV and a dominant ($BR_{b\bar b}=$100\%) annihilation branching ratio to $b\bar{b}$ \cite{abazajian, calore, daylan}, a mass of $M_{\chi} = 10$ GeV with final state $\tau^{\pm}$ \cite{abazajian, HooperGoodenough}, as well as a mixed annihilation final state with $BR_{b\bar{b}} = BR_{\tau^{\pm}} = 0.5$ and $M_{\chi} = 40$ GeV \cite{GordonMacias} (hereafter referred to as the $b\bar{b}/\tau^+\tau^-$ final state). In passing we note that (a) in the context of Majorana particle dark matter models, for example the lightest neutralino of supersymmetry, such final states are often dominant in that mass range, and (b) the spectral features of the $b\bar{b}$ is largely representative of any dark matter annihilation spectrum to strongly interacting particles (gluon or lighter quark-antiquark pairs) in the mass range under consideration.

In addition, below we also fit to the M31 data with both the mass and cross-section as free parameters, to establish the preferred mass and annihilation rate combinations that best fit the M31 emission, for the same three annihilation final states listed above. While the cross-section allows us to normalize the predicted emission to the Fermi data, adjusting the mass allows us to shift the peak of the spectrum to better fit the data. In doing so, we explore the compatibility between the particle models that fit the M31 excess with models that fit the GCE.

\begin{table*}[tbph]
\centering
\def\arraystretch{1.5}
\begin{tabular}{cccccccc}
	\hline\hline
	Free Parameters &Channel& $M_{\chi}$ (GeV)\:\:&\multicolumn{3}{c}{$\avg{\sigma v}$ (10$^{-26}$ cm$^{3}$s$^{-1}$)}&$\:\:\chi^2_{min}$&\:\:$p$-value \\ \cline{4-6}
&&&$\:\:\gamma=1.00\:\:$&$\:\:\gamma=1.25\:\:$&$\:\:\gamma=1.50\:\:$&&\\
	\hline
	\multirow{6}{*}{$\avg{\sigma v}$}      & $b\bar{b}$  &  40 & 33.19&  $7.35$  &0.56&  25.35  &  \:\:$2.9\times 10^{-4}$ \\
							&$\tau^+\tau^-$     &  40  &39.17&  $8.69$ & 0.66&  62.72  &  \:\:$< 10^{-5}$ \\
							&$b\bar{b}/\tau^+\tau^-$  &  40  &47.86  &$10.64$ &0.81 &  31.48  &  \:\:$2.1\times 10^{-5}$ \\
							&$b\bar{b}$  &  10  & 10.84& $2.41$  &0.18&  6.06  &  \:\:$0.42$ \\
							&$\tau^+\tau^-$    &  10  &9.83&  $2.18$  &0.17&  44.24  &  \:\:$< 10^{-5}$ \\
							&$b\bar{b}/\tau^+\tau^-$  &  10  &14.7 &$3.04$  &0.23&  10.01  &  \:\:$0.12$ \\
	 \hline
	\multirow{2}{*}{$M_{\chi},\avg{\sigma v}$}  &$b\bar{b}$  &  11.00  &11.71&  $2.60$  &0.20&  5.87  &  \:\:$0.32$ \\
													    &$b\bar{b}/\tau^+\tau^-$  &  5.80  &0.91 & $2.03$ & 0.15&  6.03  &  \:\:$0.30$ \\
	 \hline\hline
\end{tabular}
\caption{\label{tab:Fits}Parameters and results of the fitting procedure, including by column order the free parameters of each fit, the annihilation channel assumed, the DM particle mass, the cross-section for each $\gamma$ value, and the corresponding $\chi_{min}^2$ and $p$-values.}
\end{table*}

\section{\label{sec:results}Results}
\subsection{\label{sec:gamma}Compatibility with Galactic Center Excess Particle Models }

\begin{figure}[tbh!]
\centering
\includegraphics[width=0.45\textwidth]{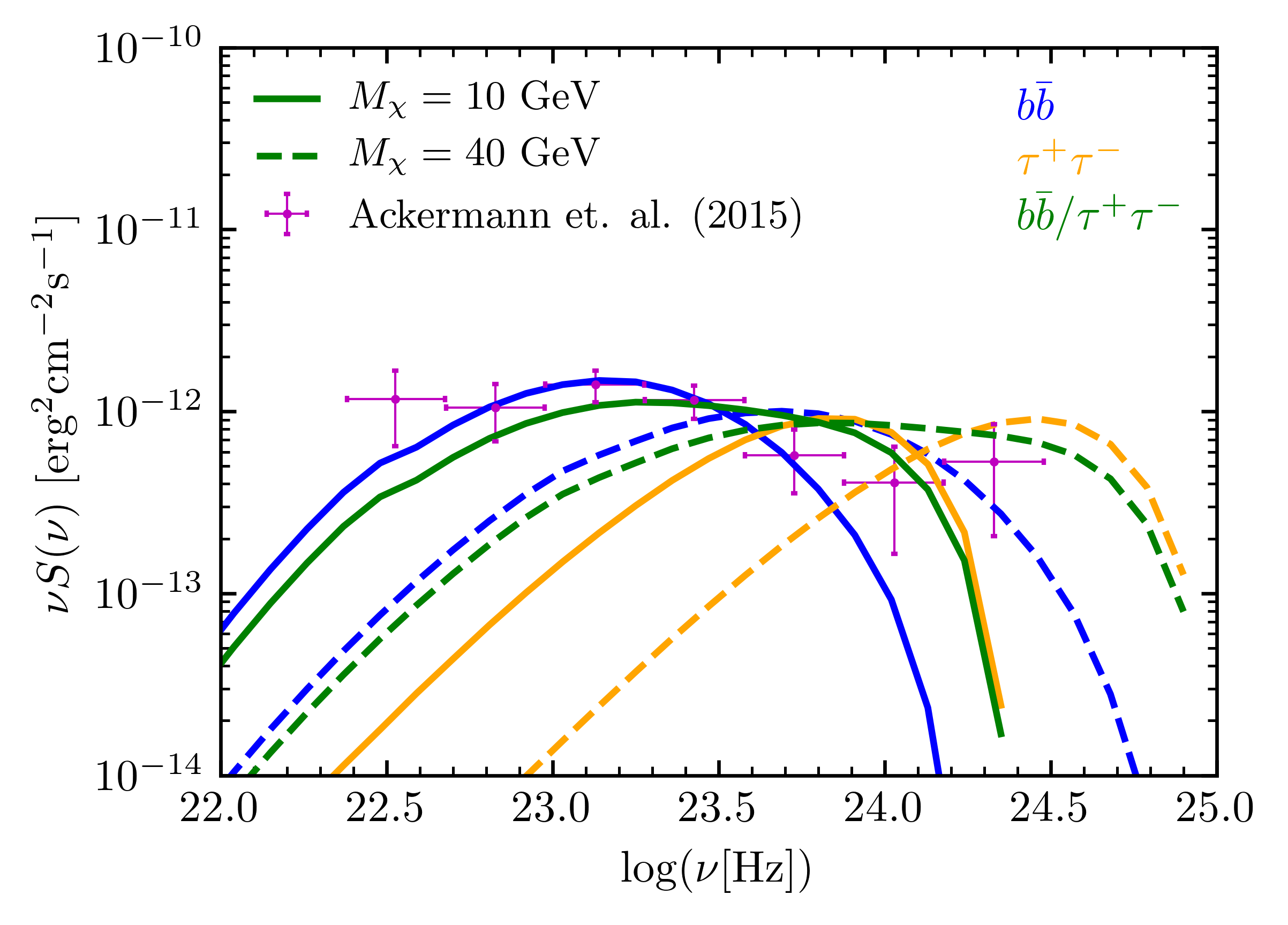}
\hfill
\caption{\label{fig:gammaSpectra}Gamma-ray SED using best-fit cross-sections. with fixed mass values of $M_{\chi} = 10$ GeV (solid) and $M_{\chi} = 40$ GeV (dashed) for three final states, $b\bar{b}$ (blue), $\tau^+\tau^-$ (orange), and $b\bar{b}/\tau^+\tau^-$ (green).}
\end{figure}

For the initial comparison with GCE particle models we choose the DM particle mass and annihilation channel to be fixed and adjust the normalization. We consider particle models with masses of 10 and 40 GeV, annihilation final states $b\bar{b}$, $\tau^+\tau^-$, and $b\bar{b}/\tau^+\tau^-$, and cross-sections in the range $\sim 10^{-30}-10^{-20}$ cm$^{3}$s$^{-1}$. Cross-sections were determined by minimizing $\chi^2$, and the results are reported in table \ref{tab:Fits} along with the corresponding $\chi_{min}^2$ and $p$-values. The normalized gamma-ray spectra in an ROI of 5 kpc, corresponding to the region where the observed excess is concentrated, for each particle mass and annihilation state are shown in figure \ref{fig:gammaSpectra} along with the Fermi M31 data \cite{FermiM31}. The best fitting particle model for the masses considered is given by the $b\bar{b}$ model at 10 GeV, followed by the mixed $b\bar{b}/\tau^+\tau^-$ final state at 10 GeV. The pure $\tau^+\tau^-$ annihilation channel at 10 GeV and 40 GeV as well as the $b\bar{b}$ and $b\bar{b}/\tau^+\tau^-$ 40 GeV particle models have harder spectra that do not fit the Fermi M31 data well. 

\begin{figure}[tbh!]
\centering
\includegraphics[width=0.45\textwidth]{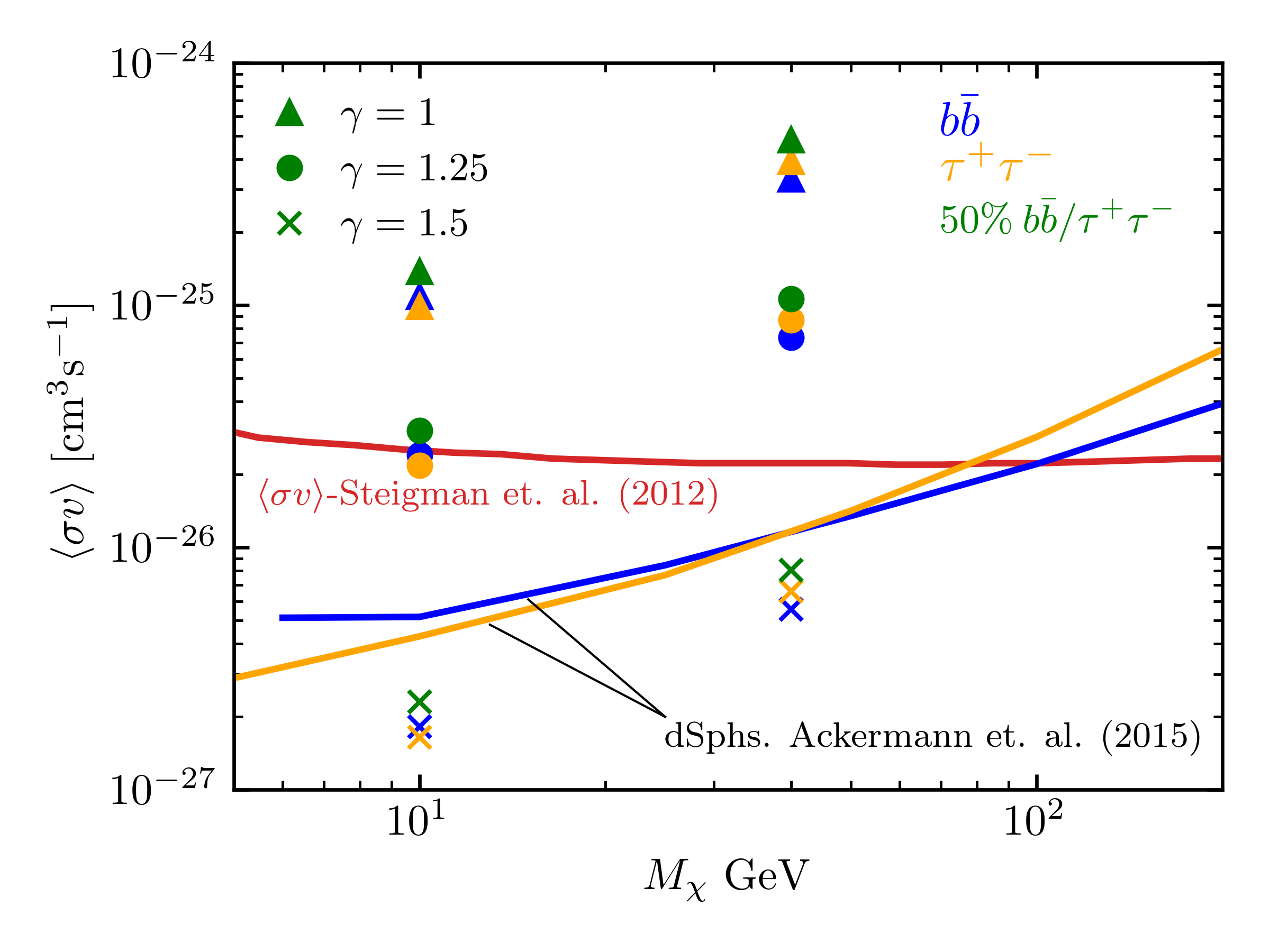}
\hfill
\caption{The normalized cross-sections are shown for each mass considered in comparison to the Fermi dSphs constraints for multiple $\gamma$ values and annihilation channels.}
\label{fig:ExCurvePoints}
\end{figure}

The spectra in figure \ref{fig:gammaSpectra} assume a NFW parameter of $\gamma = 1.25$ in accordance with the discussion in section \ref{sec:dmProfile}. However, the actual steepness of the inner profile (i.e. $\gamma$) is uncertain, so we show in figure \ref{fig:ExCurvePoints} the best-fitting cross-section of the dark matter particle masses and final states under consideration for a variety of $\gamma$ values. The resulting particle models are compared with Fermi gamma-ray constraints from observations of dSphs \cite{Ackermann2015}.

For all masses and annihilation channels, the shallowest DM profile ($\gamma$ = 1) conflicts with the Fermi dSphs results, requiring cross-sections well above the reported constraints. At $\gamma=1.25$, the required cross-sections for the 40 GeV particles are still almost an order of magnitude higher than the maximal annihilation cross-sections allowed by Fermi dSph constraints. For the 10 GeV models at $\gamma=1.25$, all of the cross-sections are concentrated at $\sim 2-3 \times 10^{-26}$ cm$^{3}$s$^{-1}$, or right around the thermal relic cross-section \cite{steigman}. In the case of the $\tau^+\tau^-$ annihilation channel, this is in good agreement with best-fit results of GCE analysis, although conflicts in the case of $b\bar{b}$ final states, since higher masses ($\sim 40$ GeV) are favored for $b\bar{b}$ in GCE particle models. As we steepen the profile to $\gamma = 1.5$, we find that the necessary cross-sections fall below the dwarf constraints, and roughly a factor of 10 and a factor of five below the thermal-relic cross-section for the 10 GeV and 40 GeV masses respectively. Of course, a sub-thermal annihilation rate is perfectly fine from a cosmological standpoint, given for instance non-thermal production of dark matter from the decay of a heavier species in the early universe.

\subsection{\label{sec:gammaPart2}{Fitting the Mass and Cross-section to the Andromeda Gamma-ray Data}}

\begin{figure}[tbh!]
\centering
\includegraphics[width=0.45\textwidth]{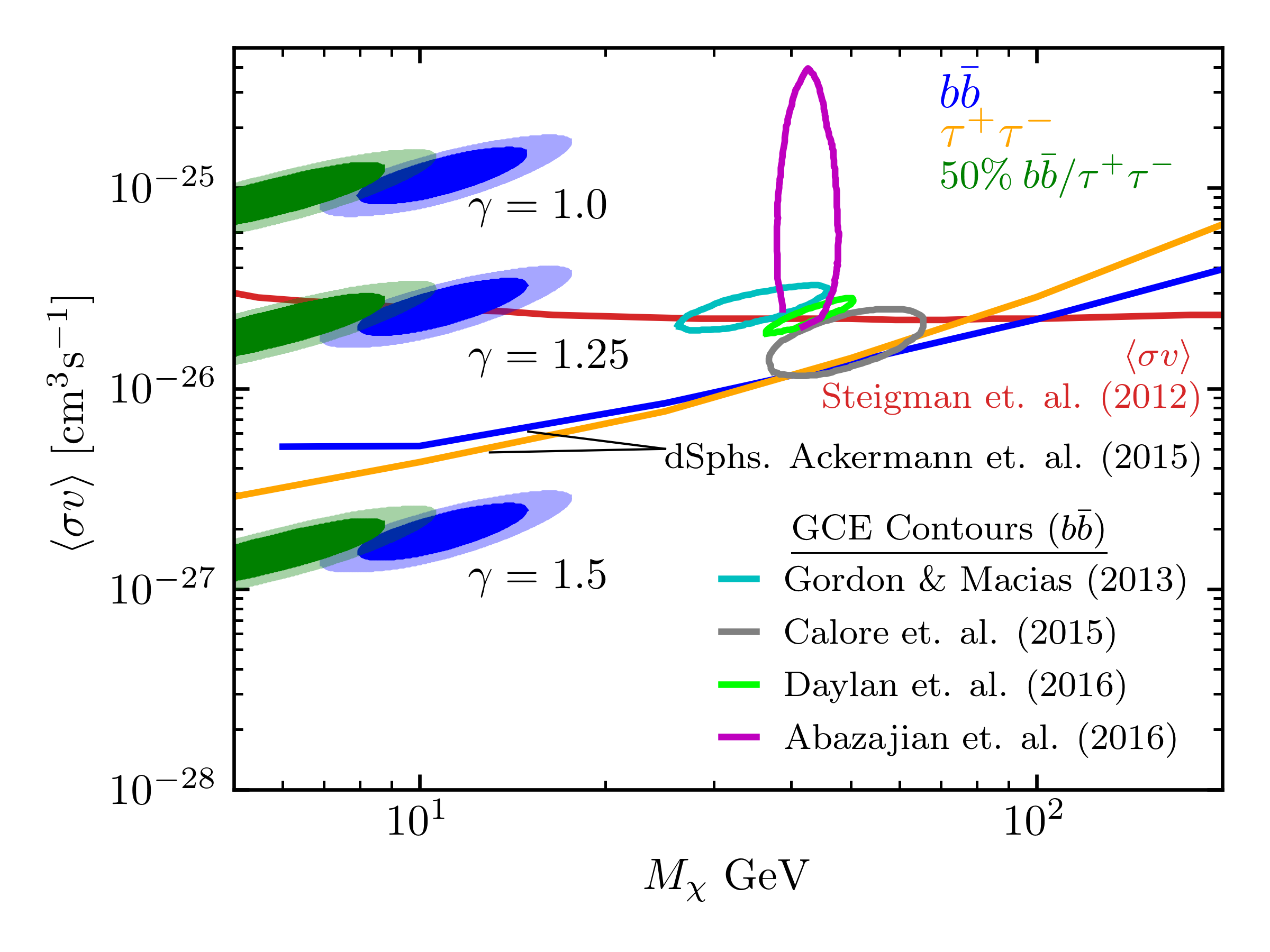}
\hfill
\caption{We show the 68\% and 95\% confidence contours of our best fit models with both $M_{\chi}$ and $\avg{\sigma v}$ as free parameters. For comparison we also show Fermi dSphs cross-section constraints, along with 95\% confidence contours for $b\bar{b}$ final states in GCE studies.}
\label{fig:ExCurveContours}
\end{figure}

Expanding on the analysis in the previous section we now allow both the mass of the dark matter as well as the cross-section to vary in order to fit the gamma-ray data and compare with GCE models. We consider masses in the range of $\sim 5-500$ GeV and cross-sections on the orders $\sim 10^{-30} -10^{-20}$ cm$^3$s$^{-1}$ and find our best fitting value in the case of $b\bar{b}$ final states with a mass of $\sim 10$ GeV as our best-fit to the Fermi M31 data. To illustrate this point quantitatively, we show the results of the fits in table \ref{tab:Fits} and show in figure \ref{fig:ExCurveContours} the 68\% and 95\% confidence levels for the $b\bar{b}$ and $b\bar{b}/\tau^+\tau^-$ final states. For a pure $\tau^+\tau^-$ final state we were unable to find a reasonable best-fit in the mass ranges considered without reaching the mass threshold for $\tau^+\tau^-$ at $\sim 1.78$ GeV.

\begin{figure}[tbh!]
\centering
\includegraphics[width=0.45\textwidth]{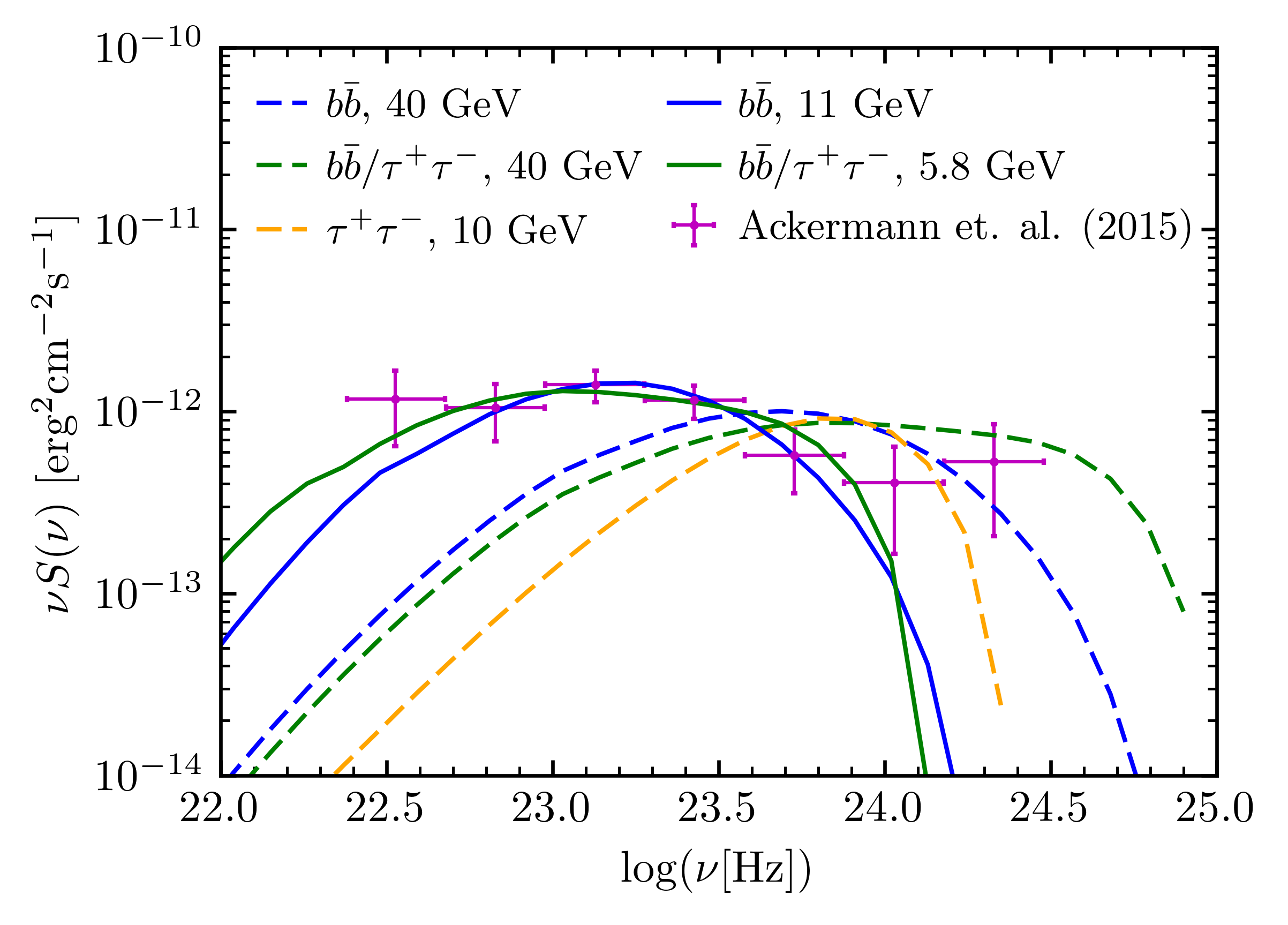}
\hfill
\caption{\label{fig:gammaSpectra2}Gamma-ray SED for the three GCE models (dashed lines) where we choose fixed masses and annihilation final states consistent with GCE models and subsequently adjust the normalizing cross-section. We additionally show the results of our best-fit models from section \ref{sec:gammaPart2} with both the mass and cross-sections as free parameters (solid lines).} 
\end{figure}

 In contrast to the previous GCE studies that found good fits for $b\bar{b}$ at $\sim 40$ GeV, we find that for all models considered a lower mass is required to fit the Fermi M31 observations. Specifically, with $\gamma = 1.25$, our best fitting model is the $b\bar{b}$ final state with $M_{\chi} = 11$ GeV and $\avg{\sigma v} = 2.60\times 10^{-26}$ cm$^{3}$ s$^{-1}$, as well as the $b\bar{b}/\tau^+\tau^-$ final state with $M_{\chi} = 5.8$ GeV and $\avg{\sigma v} = 2.03\times 10^{-26}$ cm$^{3}$ s$^{-1}$. In figure \ref{fig:gammaSpectra2}, we show the SEDs of these models in comparison to the spectra of the GCE particle models as discussed in sections \ref{sec:particle} and \ref{sec:gamma}, and with $\gamma=1.25$ as in figure \ref{fig:gammaSpectra}.
\subsection{\label{sec:radio}Comparison to Radio Data }
\begin{figure}[tbh!]
\centering
\includegraphics[width=0.45\textwidth]{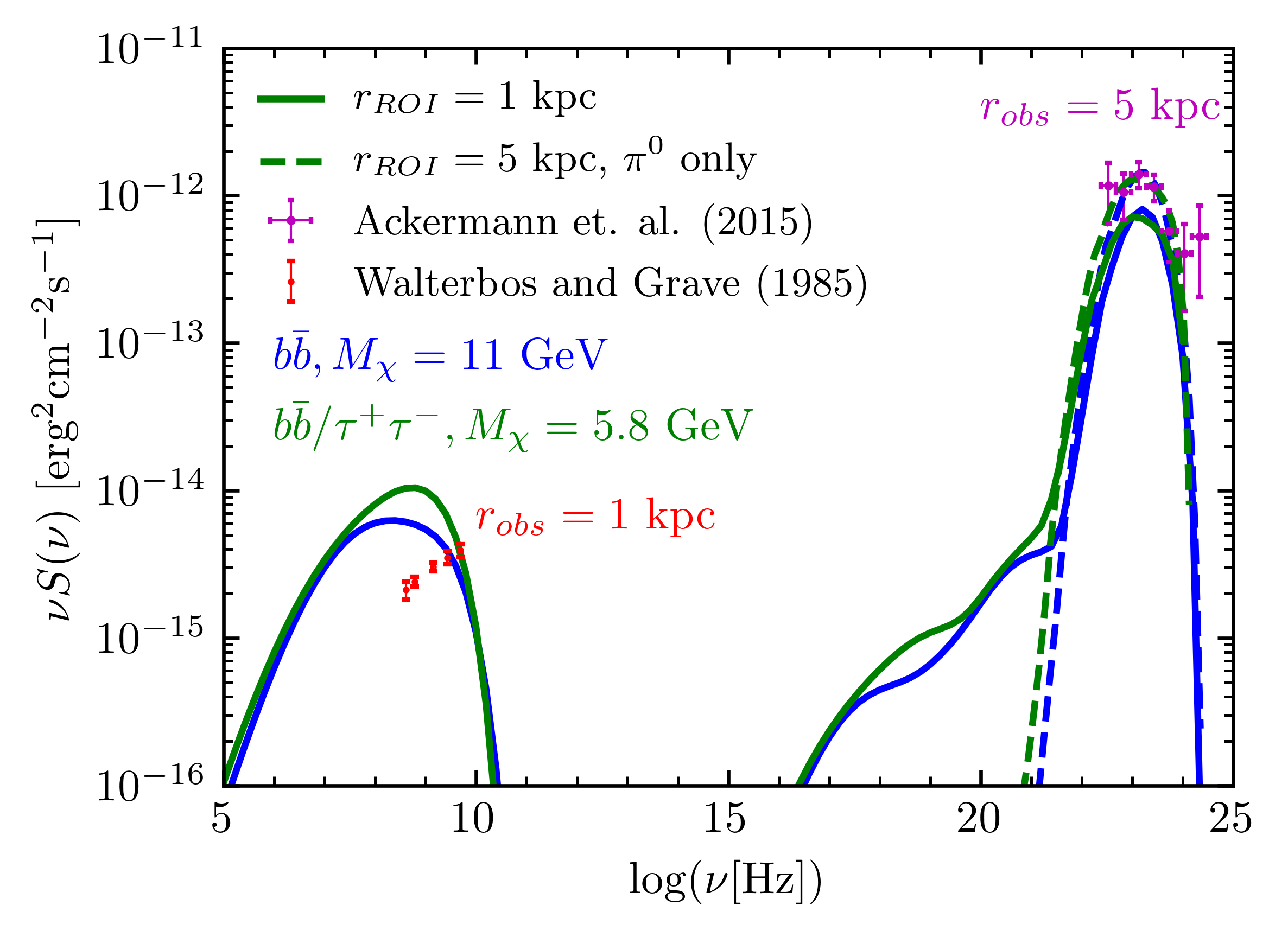}
\hfill
\caption{Multi-wavelength SED of our two best-fit models within a 1 kpc ROI compared to radio data \cite{WalterbosGrave}. Also shown is the Fermi data \cite{FermiM31} and our predicted gamma-ray emission (dashed) within a 5 kpc region.}
\label{fig:bb_btSED}
\end{figure}

The annihilation of WIMP dark matter particles is expected to produce not only gamma-rays through neutral pion decay, but also an abundance of charged electron/positron pairs, which in turn are expected to produce radio emissions through synchrotron radiation. Thus, any DM particle model that is purported to explain the gamma-ray excess in M31 should also be compatible with radio observations under reasonable assumptions for magnetic field and diffusion models. The focus of radio studies in the literature has largely been on studying emissions in larger regions of M31 out to radii of $\sim 16$ kpc \cite{RadioData, M31_350, BeckBerkWiel, BeckBerkHoernes, BeckGrave}, or in the very central $\sim 1$ kpc region \cite{WalterbosGrave, BerkWielBeck}. To account for this we use the particle models consistent with the Fermi emission in a 5 kpc radius, then predict the radio emissions in a 1 kpc radius in order to compare with observational radio data. When extrapolating from 5 kpc to 1 kpc we assume an inner DM density profile with $\gamma = 1.25$. Additionally, we adopt diffusion and magnetic field models as described in sections \ref{sec:diffusion} and \ref{sec:Bfield} respectively of $D(E) = D_0 E^{\delta}$ with $D_0 =3\times 10^{28}$ cm$^{2}$ s$^{-1}$ and $\delta =0.3$ and $B(r) = B_0 e^{-r/r_c} + B_{const}$ with $B_0=10\:\mu$G and $B_{const} =5\:\mu$G (see also table \ref{tab:Params}). Figure \ref{fig:bb_btSED} shows the multi-wavelength SED within 1 kpc for our best-fit models as determined in the previous section compared to radio data. We also include the Fermi gamma-ray data and the predicted gamma-ray emission at 5 kpc (dashed lines) for reference and to emphasize that the particle cross-section is determined by fitting to the gamma-ray data (see previous section). We note that in the mixed state scenario radio emission is predicted to be much larger than radio data in the 1 kpc region, suggesting tension with the assumption that a DM particle with this mass and annihilation state is responsible for the detected gamma-ray emissions. The $b\bar{b}$ final state model also conflicts with current observations, and predicts higher emission than observed for most data available, albeit with lower expected emissions than the mixed state model.

\begin{figure}[tbh!]
\centering
\includegraphics[width=0.45\textwidth]{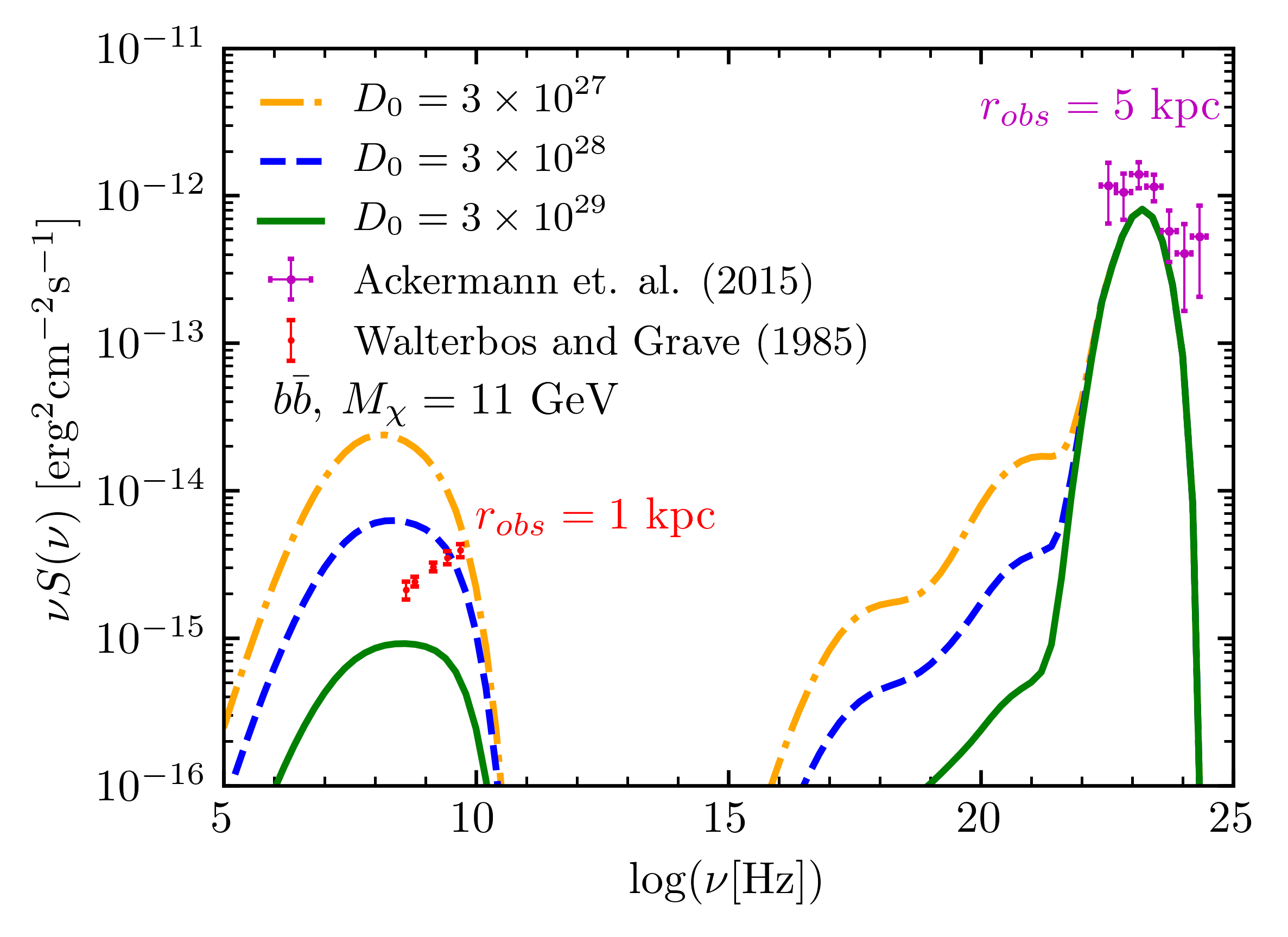}
\hfill
\caption{SED for the $b\bar{b}$ annihilation channel, $M_{\chi} = 11$ GeV, and $\avg{\sigma v} =2.6\times 10^{-26}$ cm$^3$s$^{-1}$ with multiple diffusion constant values in units of cm$^{2}$s$^{-1}$}
\label{fig:DiffSED}
\end{figure}

\begin{figure}[tbh!]
\centering
\includegraphics[width=0.45\textwidth]{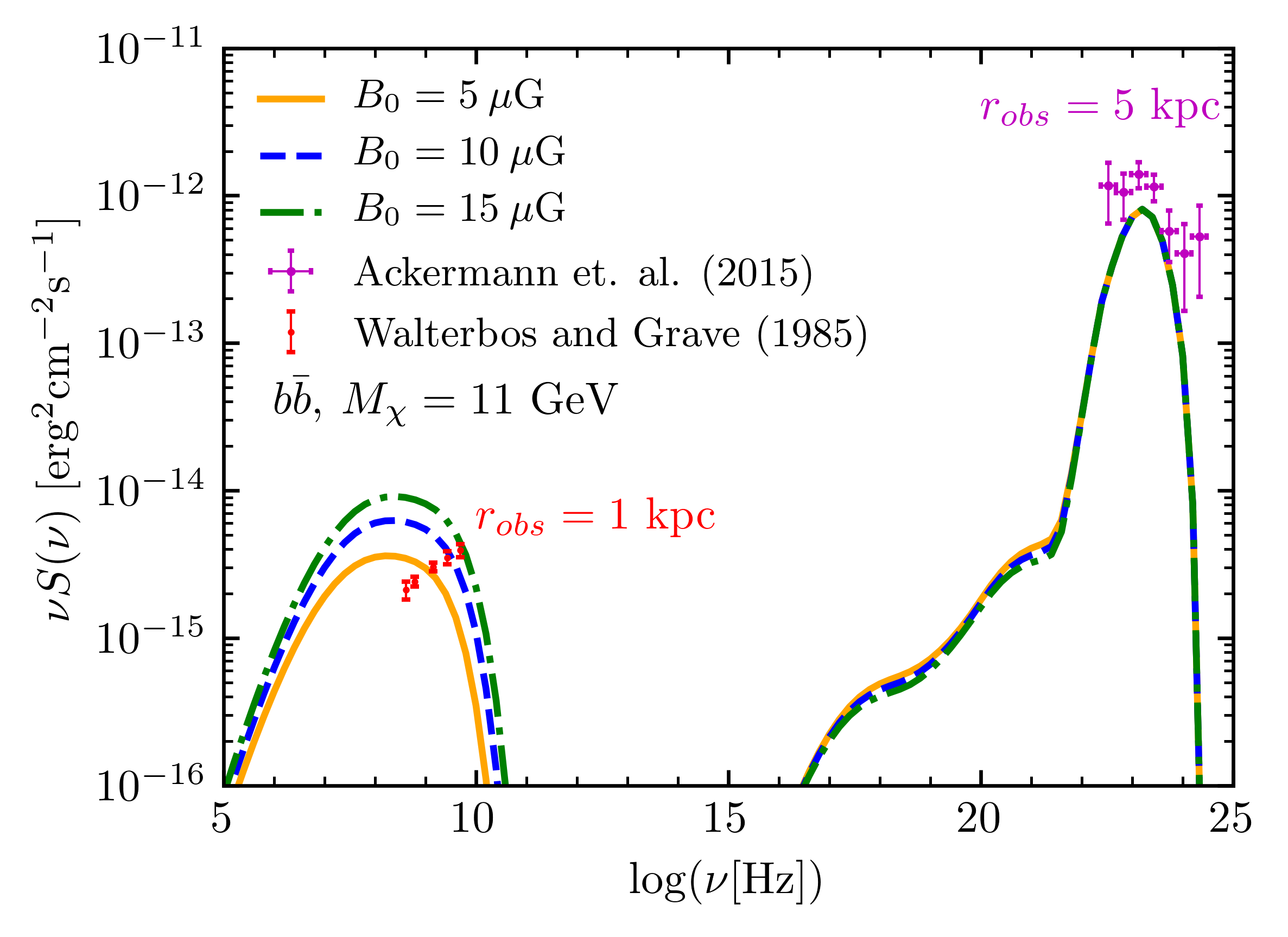}
\hfill
\caption{SED for the $b\bar{b}$ annihilation channel, $M_{\chi} = 11$ GeV, and $\avg{\sigma v} =2.6\times 10^{-26}$ cm$^3$s$^{-1}$ with multiple magnetic field strengths.}
\label{fig:BSED}
\end{figure}

We also take into consideration uncertainty in the magnetic field and the efficiency of diffusion as parameterized by the size of the diffusion coefficient. Previous studies of radio emissions due to DM annihilation in M31 have typically ignored diffusion, however in figure \ref{fig:DiffSED} we demonstrate the effect that varying the diffusion strength (over a range of low to high estimates for the Milky Way) has on the expected emissions. Additionally, in figure \ref{fig:BSED} we show the DM emission including the magnetic field uncertainty as discussed in section \ref{sec:Bfield}. Figures \ref{fig:DiffSED} and \ref{fig:BSED} emphasize the role that the uncertainties in the astrophysics of diffusion and magnetic field parameters have on our ability to make concrete statements concerning the validity of the DM explanation of the gamma-ray excess in M31. For instance, from figure \ref{fig:BSED} we see that even in the case of the lowest magnetic fields considered, the DM interpretation appears to conflict with the available data. To make dark matter compatible with the radio observations would require a magnetic field strength lower than our most conservative estimates. Diffusion however presents a much more impactful source of uncertainty, as demonstrated in figure \ref{fig:DiffSED}. While our nominal value for the diffusion constant is the most typical assumed value for the Milky Way, if we adopt a value that is at the upper limit of quoted values we see a significant decrease in the radio emission. Conversely, for decreasing the diffusion constant yields expected emission that greatly overproduces the observational data. Nevertheless, we have demonstrated that conservative estimates for these parameters predict radio emissions from DM annihilation that are in tension with current observations. 

Some other points to note include that the spectral shape of the radio emission cannot be matched by the models that fit the gamma-ray data. Due to the low masses needed to fit the gamma-rays, the synchrotron emission peaks at frequencies that are too low for the spectral shape of the predicted emission to match the observations. Additionally, in this analysis we have assumed that the radio emission observed is due entirely to dark matter annihilation. This gives a more conservative approach, since there are other astrophysical contributions to the radio emission such as synchrotron-emitting cosmic-rays that have not been taken into account.

\section{\label{sec:conclusion}Conclusion}
In this work we have examined the gamma-ray excess in M31 reported by the Fermi collaboration in the context of the multi-wavelength emissions from WIMP dark matter annihilation. We used the RX-DMFIT tool \cite{McDaniel} to predict the gamma-ray spectra from DM annihilation in M31 and to fit the expected gamma-ray signal to the Fermi data in order to develop best-fit particle models. We did this in two ways; first, we adopted the final states and DM mass values that are consistent with the GCE, and fit the cross-section to the M31 data. This allowed us to compare the necessary cross-sections assuming GCE particle models with current constraints on the DM particle. We found that particle models typically associated with the GCE do not produce spectra that provide good fits to the M31 data.

We then allowed both the cross-section and mass to be free parameters in our fit. Our best fit models in this approach were for $b\bar{b}$ final states with a mass of $M_{\chi} =11$ GeV and $\avg{\sigma v} =2.6\times 10^{-26}$ cm$^3$s$^{-1}$, consistent with other previous studies of gamma-rays in M31 \cite{LiM31}, as well as finding a reasonable fit for $b\bar{b}/\tau^+\tau^-$ final states with a mass of $M_{\chi} =5.8$ GeV and $\avg{\sigma v} =2.03\times 10^{-26}$ cm$^3$s$^{-1}$. Our analysis does not find a good fit for $\tau^+\tau^-$, although previous studies have also begun to disfavor this annihilation channel for typical GCE mass of $~7-10$ GeV (see e.g. Ref.~\cite{daylan, GordonMacias}). We noted that the M31 data tends to favor lower particle masses than the GCE data for all annihilation channels.

Finally, after establishing the class of particle dark matter models consistent with the observed gamma-ray emission, we used RX-DMFIT to calculate the expected emission due to synchrotron radiation and inverse Compton scattering of CMB and starlight photons. We compared the expected radio emission in the central regions of M31 to observational data adopting a DM particle model with $b\bar{b}$ final states with a mass of $M_{\chi} =11$ GeV and $\avg{\sigma v} =2.6\times 10^{-26}$ cm$^3$s$^{-1}$ in accordance with our best-fit to the gamma-ray data. In this scenario we found that the expected emissions tend to overproduce in the observed radio emission for conservative estimates of the magnetic field ($B_0 =10 \mu$G) and diffusion constant ($D_0 = 3\times 10^{28}$ cm$^2$ s$^{-1}$). However, our study shows that very efficient diffusion in M31, to levels around one of order of magnitude larger than in the Milky Way, or highly suppressed average magnetic fields, could reconcile the relatively dim radio emission observed from the innermost 1 kpc of M31 with the expected bright radio emission from secondary electrons and positrons produced by dark matter annihilation.

\begin{acknowledgments}
This study is based on work supported by the National Science Foundation under Grant No. 1517545. S.P. is partly supported by the US Department of Energy, grant number DE-SC0010107.
\end{acknowledgments}
\bibliographystyle{unsrtnat}
\bibliography{M31ref2}

\begin{thebibliography}{65}
\providecommand{\natexlab}[1]{#1}
\providecommand{\url}[1]{\texttt{#1}}
\expandafter\ifx\csname urlstyle\endcsname\relax
  \providecommand{\doi}[1]{doi: #1}\else
  \providecommand{\doi}{doi: \begingroup \urlstyle{rm}\Url}\fi

\bibitem[{Bertone} et~al.(2005){Bertone}, {Hooper}, and {Silk}]{bertone}
G.~{Bertone}, D.~{Hooper}, and J.~{Silk}.
\newblock {Particle dark matter: evidence, candidates and constraints}.
\newblock \emph{\physrep}, 405:\penalty0 279--390, January 2005.
\newblock \doi{10.1016/j.physrep.2004.08.031}.

\bibitem[{Jungman} et~al.(1996){Jungman}, {Kamionkowski}, and
  {Griest}]{jungman}
G.~{Jungman}, M.~{Kamionkowski}, and K.~{Griest}.
\newblock {Supersymmetric dark matter}.
\newblock \emph{\physrep}, 267:\penalty0 195--373, March 1996.
\newblock \doi{10.1016/0370-1573(95)00058-5}.

\bibitem[{Bergstr{\"o}m}(2012)]{bergstrom}
L.~{Bergstr{\"o}m}.
\newblock {Dark matter evidence, particle physics candidates and detection
  methods}.
\newblock \emph{Annalen der Physik}, 524:\penalty0 479--496, October 2012.
\newblock \doi{10.1002/andp.201200116}.

\bibitem[{Storm} et~al.(2013){Storm}, {Jeltema}, {Profumo}, and
  {Rudnick}]{storm}
E.~{Storm}, T.~E. {Jeltema}, S.~{Profumo}, and L.~{Rudnick}.
\newblock {Constraints on Dark Matter Annihilation in Clusters of Galaxies from
  Diffuse Radio Emission}.
\newblock \emph{\apj}, 768:\penalty0 106, May 2013.
\newblock \doi{10.1088/0004-637X/768/2/106}.

\bibitem[{Storm} et~al.(2016){Storm}, {Jeltema}, {Splettstoesser}, and
  {Profumo}]{storm16}
E.~{Storm}, T.~E. {Jeltema}, M.~{Splettstoesser}, and S.~{Profumo}.
\newblock {Synchrotron Emission from Dark Matter Annihilation: Predictions for
  Constraints from Non-detections of Galaxy Clusters with New Radio Surveys}.
\newblock \emph{ArXiv e-prints}, July 2016.

\bibitem[{Egorov} and {Pierpaoli}(2013)]{M31USC}
A.~E. {Egorov} and E.~{Pierpaoli}.
\newblock {Constraints on dark matter annihilation by radio observations of
  M31}.
\newblock \emph{\prd}, 88\penalty0 (2):\penalty0 023504, July 2013.
\newblock \doi{10.1103/PhysRevD.88.023504}.

\bibitem[{Chan}(2016)]{chan}
M.~H. {Chan}.
\newblock {Revisiting the constraints on annihilating dark matter by the radio
  observational data of M31}.
\newblock \emph{\prd}, 94\penalty0 (2):\penalty0 023507, July 2016.
\newblock \doi{10.1103/PhysRevD.94.023507}.

\bibitem[{Laha} et~al.(2013){Laha}, {Ng}, {Dasgupta}, and {Horiuchi}]{laha}
R.~{Laha}, K.~C.~Y. {Ng}, B.~{Dasgupta}, and S.~{Horiuchi}.
\newblock {Galactic Center radio constraints on gamma-ray lines from dark
  matter annihilation}.
\newblock \emph{\prd}, 87\penalty0 (4):\penalty0 043516, February 2013.
\newblock \doi{10.1103/PhysRevD.87.043516}.

\bibitem[{Regis} and {Ullio}(2008)]{RegisUllio}
M.~{Regis} and P.~{Ullio}.
\newblock {Multiwavelength signals of dark matter annihilations at the Galactic
  center}.
\newblock \emph{\prd}, 78\penalty0 (4):\penalty0 043505, August 2008.
\newblock \doi{10.1103/PhysRevD.78.043505}.

\bibitem[{Natarajan} et~al.(2013){Natarajan}, {Peterson}, {Voytek}, {Spekkens},
  {Mason}, {Aguirre}, and {Willman}]{Natarajan}
A.~{Natarajan}, J.~B. {Peterson}, T.~C. {Voytek}, K.~{Spekkens}, B.~{Mason},
  J.~{Aguirre}, and B.~{Willman}.
\newblock {Bounds on dark matter properties from radio observations of Ursa
  Major II using the Green Bank Telescope}.
\newblock \emph{\prd}, 88\penalty0 (8):\penalty0 083535, October 2013.
\newblock \doi{10.1103/PhysRevD.88.083535}.

\bibitem[{Spekkens} et~al.(2013){Spekkens}, {Mason}, {Aguirre}, and
  {Nhan}]{spekkens}
K.~{Spekkens}, B.~S. {Mason}, J.~E. {Aguirre}, and B.~{Nhan}.
\newblock {A Deep Search for Extended Radio Continuum Emission from Dwarf
  Spheroidal Galaxies: Implications for Particle Dark Matter}.
\newblock \emph{\apj}, 773:\penalty0 61, August 2013.
\newblock \doi{10.1088/0004-637X/773/1/61}.

\bibitem[{Natarajan} et~al.(2015){Natarajan}, {Aguirre}, {Spekkens}, and
  {Mason}]{natarajanSegue}
A.~{Natarajan}, J.~E. {Aguirre}, K.~{Spekkens}, and B.~S. {Mason}.
\newblock {Green Bank Telescope Constraints on Dark Matter Annihilation in
  Segue I}.
\newblock \emph{ArXiv e-prints}, July 2015.

\bibitem[{Colafrancesco} et~al.(2007){Colafrancesco}, {Profumo}, and
  {Ullio}]{colaDraco}
S.~{Colafrancesco}, S.~{Profumo}, and P.~{Ullio}.
\newblock {Detecting dark matter WIMPs in the Draco dwarf: A multiwavelength
  perspective}.
\newblock \emph{\prd}, 75\penalty0 (2):\penalty0 023513, January 2007.
\newblock \doi{10.1103/PhysRevD.75.023513}.

\bibitem[{Jeltema} and {Profumo}(2012)]{JP2012}
T.~E. {Jeltema} and S.~{Profumo}.
\newblock {Dark matter detection with hard X-ray telescopes}.
\newblock \emph{\mnras}, 421:\penalty0 1215--1221, April 2012.
\newblock \doi{10.1111/j.1365-2966.2011.20382.x}.

\bibitem[{Jeltema} and {Profumo}(2008)]{TeslaXray}
T.~E. {Jeltema} and S.~{Profumo}.
\newblock {Searching for Dark Matter with X-Ray Observations of Local Dwarf
  Galaxies}.
\newblock \emph{\apj}, 686:\penalty0 1045-1055, October 2008.
\newblock \doi{10.1086/591495}.

\bibitem[{Goodenough} and {Hooper}(2009)]{goodenough}
L.~{Goodenough} and D.~{Hooper}.
\newblock {Possible Evidence For Dark Matter Annihilation In The Inner Milky
  Way From The Fermi Gamma Ray Space Telescope}.
\newblock \emph{ArXiv e-prints}, October 2009.

\bibitem[{Abazajian} et~al.(2014){Abazajian}, {Canac}, {Horiuchi}, and
  {Kaplinghat}]{abazajian2014}
K.~N. {Abazajian}, N.~{Canac}, S.~{Horiuchi}, and M.~{Kaplinghat}.
\newblock {Astrophysical and dark matter interpretations of extended gamma-ray
  emission from the Galactic Center}.
\newblock \emph{\prd}, 90\penalty0 (2):\penalty0 023526, July 2014.
\newblock \doi{10.1103/PhysRevD.90.023526}.

\bibitem[{Calore} et~al.(2015){Calore}, {Cholis}, {McCabe}, and
  {Weniger}]{calore}
F.~{Calore}, I.~{Cholis}, C.~{McCabe}, and C.~{Weniger}.
\newblock {A tale of tails: Dark matter interpretations of the Fermi GeV excess
  in light of background model systematics}.
\newblock \emph{\prd}, 91\penalty0 (6):\penalty0 063003, March 2015.
\newblock \doi{10.1103/PhysRevD.91.063003}.

\bibitem[{Daylan} et~al.(2016){Daylan}, {Finkbeiner}, {Hooper}, {Linden},
  {Portillo}, {Rodd}, and {Slatyer}]{daylan}
T.~{Daylan}, D.~P. {Finkbeiner}, D.~{Hooper}, T.~{Linden}, S.~K.~N. {Portillo},
  N.~L. {Rodd}, and T.~R. {Slatyer}.
\newblock {The characterization of the gamma-ray signal from the central Milky
  Way: A case for annihilating dark matter}.
\newblock \emph{Physics of the Dark Universe}, 12:\penalty0 1--23, June 2016.
\newblock \doi{10.1016/j.dark.2015.12.005}.

\bibitem[Gordon and Mac\'{\i}as(2013)]{GordonMacias}
Chris Gordon and Oscar Mac\'{\i}as.
\newblock Dark matter and pulsar model constraints from galactic center
  fermi-lat gamma-ray observations.
\newblock \emph{Phys. Rev. D}, 88:\penalty0 083521, Oct 2013.
\newblock \doi{10.1103/PhysRevD.88.083521}.
\newblock URL \url{https://link.aps.org/doi/10.1103/PhysRevD.88.083521}.

\bibitem[{Hooper} and {Goodenough}(2011)]{HooperGoodenough}
D.~{Hooper} and L.~{Goodenough}.
\newblock {Dark matter annihilation in the Galactic Center as seen by the Fermi
  Gamma Ray Space Telescope}.
\newblock \emph{Physics Letters B}, 697:\penalty0 412--428, March 2011.
\newblock \doi{10.1016/j.physletb.2011.02.029}.

\bibitem[{Ackermann} et~al.(2017{\natexlab{a}})]{FermiReview}
M.~{Ackermann} et~al.
\newblock {The Fermi Galactic Center GeV Excess and Implications for Dark
  Matter}.
\newblock \emph{\apj}, 840:\penalty0 43, May 2017{\natexlab{a}}.
\newblock \doi{10.3847/1538-4357/aa6cab}.

\bibitem[{Eckner} et~al.(2017){Eckner}, {Hou}, {Serpico}, {Winter},
  {Zaharijas}, {Martin}, {di Mauro}, {Mirabal}, {Petrovic}, {Prodanovic}, and
  {Vandenbrouck}]{EcknerMSP}
C.~{Eckner}, X.~{Hou}, P.~D. {Serpico}, M.~{Winter}, G.~{Zaharijas},
  P.~{Martin}, M.~{di Mauro}, N.~{Mirabal}, J.~{Petrovic}, T.~{Prodanovic}, and
  J.~{Vandenbrouck}.
\newblock {Millisecond pulsar origin of the Galactic center excess and extended
  gamma-ray emission from Andromeda - a closer look}.
\newblock \emph{ArXiv e-prints}, November 2017.

\bibitem[{Bartels} et~al.(2016){Bartels}, {Krishnamurthy}, and
  {Weniger}]{bartels}
R.~{Bartels}, S.~{Krishnamurthy}, and C.~{Weniger}.
\newblock {Strong Support for the Millisecond Pulsar Origin of the Galactic
  Center GeV Excess}.
\newblock \emph{Physical Review Letters}, 116\penalty0 (5):\penalty0 051102,
  February 2016.
\newblock \doi{10.1103/PhysRevLett.116.051102}.

\bibitem[{Brandt} and {Kocsis}(2015)]{brandt}
T.~D. {Brandt} and B.~{Kocsis}.
\newblock {Disrupted Globular Clusters Can Explain the Galactic Center
  Gamma-Ray Excess}.
\newblock \emph{\apj}, 812:\penalty0 15, October 2015.
\newblock \doi{10.1088/0004-637X/812/1/15}.

\bibitem[{Carlson} et~al.(2016){Carlson}, {Profumo}, and {Linden}]{carlson}
E.~{Carlson}, S.~{Profumo}, and T.~{Linden}.
\newblock {Cosmic-Ray Injection from Star-Forming Regions}.
\newblock \emph{Physical Review Letters}, 117\penalty0 (11):\penalty0 111101,
  September 2016.
\newblock \doi{10.1103/PhysRevLett.117.111101}.

\bibitem[{Gaggero} et~al.(2015){Gaggero}, {Taoso}, {Urbano}, {Valli}, and
  {Ullio}]{gaggero}
D.~{Gaggero}, M.~{Taoso}, A.~{Urbano}, M.~{Valli}, and P.~{Ullio}.
\newblock {Towards a realistic astrophysical interpretation of the gamma-ray
  Galactic center excess}.
\newblock \emph{\jcap}, 12:\penalty0 056, December 2015.
\newblock \doi{10.1088/1475-7516/2015/12/056}.

\bibitem[{Cholis} et~al.(2015){Cholis}, {Evoli}, {Calore}, {Linden}, {Weniger},
  and {Hooper}]{cholis}
I.~{Cholis}, C.~{Evoli}, F.~{Calore}, T.~{Linden}, C.~{Weniger}, and
  D.~{Hooper}.
\newblock {The Galactic Center GeV excess from a series of leptonic cosmic-ray
  outbursts}.
\newblock \emph{\jcap}, 12:\penalty0 005, December 2015.
\newblock \doi{10.1088/1475-7516/2015/12/005}.

\bibitem[{Ackermann} et~al.(2017{\natexlab{b}})]{FermiM31}
M.~{Ackermann} et~al.
\newblock {Observations of M31 and M33 with the Fermi Large Area Telescope: A
  Galactic Center Excess in Andromeda?}
\newblock \emph{\apj}, 836:\penalty0 208, February 2017{\natexlab{b}}.
\newblock \doi{10.3847/1538-4357/aa5c3d}.

\bibitem[{McDaniel} et~al.(2017){McDaniel}, {Jeltema}, {Profumo}, and
  {Storm}]{McDaniel}
A.~{McDaniel}, T.~{Jeltema}, S.~{Profumo}, and E.~{Storm}.
\newblock {Multiwavelength analysis of dark matter annihilation and RX-DMFIT}.
\newblock \emph{\jcap}, 9:\penalty0 027, September 2017.
\newblock \doi{10.1088/1475-7516/2017/09/027}.

\bibitem[{Beck} and {Colafrancesco}(2017)]{colaM31}
G.~{Beck} and S.~{Colafrancesco}.
\newblock {A multi-frequency analysis of possible dark matter contributions to
  M31 gamma-ray emissions}.
\newblock \emph{\jcap}, 10:\penalty0 007, October 2017.
\newblock \doi{10.1088/1475-7516/2017/10/007}.

\bibitem[Longair(2011)]{longair}
Malcolm~S. Longair.
\newblock \emph{High Energy Astrophysics}.
\newblock Cambridge University Press, New York, 3 edition, 2011.
\newblock ISBN 978-0-521-75618-1.

\bibitem[{Colafrancesco} et~al.(2006){Colafrancesco}, {Profumo}, and
  {Ullio}]{cola}
S.~{Colafrancesco}, S.~{Profumo}, and P.~{Ullio}.
\newblock {Multi-frequency analysis of neutralino dark matter annihilations in
  the Coma cluster}.
\newblock \emph{\aap}, 455:\penalty0 21--43, August 2006.
\newblock \doi{10.1051/0004-6361:20053887}.

\bibitem[{Profumo} and {Ullio}(2010)]{ProfumoUllio}
S.~{Profumo} and P.~{Ullio}.
\newblock {Multi-wavelength Searches for Particle Dark Matter}.
\newblock \emph{ArXiv e-prints}, January 2010.

\bibitem[{Porter} et~al.(2008){Porter}, {Moskalenko}, {Strong}, {Orlando}, and
  {Bouchet}]{porter}
T.~A. {Porter}, I.~V. {Moskalenko}, A.~W. {Strong}, E.~{Orlando}, and
  L.~{Bouchet}.
\newblock {Inverse Compton Origin of the Hard X-Ray and Soft Gamma-Ray Emission
  from the Galactic Ridge}.
\newblock \emph{\apj}, 682:\penalty0 400-407, July 2008.
\newblock \doi{10.1086/589615}.

\bibitem[{Ginzburg} and {Syrovatskii}(1964)]{ginzburg}
V.~L. {Ginzburg} and S.~I. {Syrovatskii}.
\newblock \emph{{The Origin of Cosmic Rays}}.
\newblock 1964.

\bibitem[{Strong} et~al.(2007){Strong}, {Moskalenko}, and {Ptuskin}]{SMP}
A.~W. {Strong}, I.~V. {Moskalenko}, and V.~S. {Ptuskin}.
\newblock {Cosmic-Ray Propagation and Interactions in the Galaxy}.
\newblock \emph{Annual Review of Nuclear and Particle Science}, 57:\penalty0
  285--327, November 2007.
\newblock \doi{10.1146/annurev.nucl.57.090506.123011}.

\bibitem[{Vladimirov} et~al.(2012){Vladimirov}, {J{\'o}hannesson},
  {Moskalenko}, and {Porter}]{vladimirov}
A.~E. {Vladimirov}, G.~{J{\'o}hannesson}, I.~V. {Moskalenko}, and T.~A.
  {Porter}.
\newblock {Testing the Origin of High-energy Cosmic Rays}.
\newblock \emph{\apj}, 752:\penalty0 68, June 2012.
\newblock \doi{10.1088/0004-637X/752/1/68}.

\bibitem[{Baltz} and {Edsj{\"o}}(1998)]{BaltzEdsjo}
E.~A. {Baltz} and J.~{Edsj{\"o}}.
\newblock {Positron propagation and fluxes from neutralino annihilation in the
  halo}.
\newblock \emph{\prd}, 59\penalty0 (2):\penalty0 023511, January 1998.
\newblock \doi{10.1103/PhysRevD.59.023511}.

\bibitem[{Webber} et~al.(1992){Webber}, {Lee}, and {Gupta}]{WLG}
W.~R. {Webber}, M.~A. {Lee}, and M.~{Gupta}.
\newblock {Propagation of cosmic-ray nuclei in a diffusing galaxy with
  convective halo and thin matter disk}.
\newblock \emph{\apj}, 390:\penalty0 96--104, May 1992.
\newblock \doi{10.1086/171262}.

\bibitem[{Crocker} et~al.(2011){Crocker}, {Jones}, {Aharonian}, {Law}, {Melia},
  {Oka}, and {Ott}]{winds}
R.~M. {Crocker}, D.~I. {Jones}, F.~{Aharonian}, C.~J. {Law}, F.~{Melia},
  T.~{Oka}, and J.~{Ott}.
\newblock {Wild at Heart: the particle astrophysics of the Galactic Centre}.
\newblock \emph{\mnras}, 413:\penalty0 763--788, May 2011.
\newblock \doi{10.1111/j.1365-2966.2010.18170.x}.

\bibitem[{Gie{\ss}{\"u}bel} and {Beck}(2014)]{giessubel}
R.~{Gie{\ss}{\"u}bel} and R.~{Beck}.
\newblock {The magnetic field structure of the central region in M 31}.
\newblock \emph{\aap}, 571:\penalty0 A61, November 2014.
\newblock \doi{10.1051/0004-6361/201323211}.

\bibitem[{Hoernes} et~al.(1998){Hoernes}, {Beck}, and {Berkhuijsen}]{hoernes}
P.~{Hoernes}, R.~{Beck}, and E.~M. {Berkhuijsen}.
\newblock {Properties of synchrotron emission and magnetic fields in the
  central region of M31}.
\newblock In Y.~{Sofue}, editor, \emph{The Central Regions of the Galaxy and
  Galaxies}, volume 184 of \emph{IAU Symposium}, page 351, 1998.

\bibitem[{Fletcher} et~al.(2004){Fletcher}, {Berkhuijsen}, {Beck}, and
  {Shukurov}]{fletcher}
A.~{Fletcher}, E.~M. {Berkhuijsen}, R.~{Beck}, and A.~{Shukurov}.
\newblock {The magnetic field of M 31 from multi-wavelength radio polarization
  observations}.
\newblock \emph{\aap}, 414:\penalty0 53--67, January 2004.
\newblock \doi{10.1051/0004-6361:20034133}.

\bibitem[{Navarro} et~al.(1996){Navarro}, {Frenk}, and {White}]{NFW96}
J.~F. {Navarro}, C.~S. {Frenk}, and S.~D.~M. {White}.
\newblock {The Structure of Cold Dark Matter Halos}.
\newblock \emph{\apj}, 462:\penalty0 563, May 1996.
\newblock \doi{10.1086/177173}.

\bibitem[{Navarro} et~al.(1997){Navarro}, {Frenk}, and {White}]{NFW97}
J.~F. {Navarro}, C.~S. {Frenk}, and S.~D.~M. {White}.
\newblock {A Universal Density Profile from Hierarchical Clustering}.
\newblock \emph{\apj}, 490:\penalty0 493--508, December 1997.
\newblock \doi{10.1086/304888}.

\bibitem[{Tamm} et~al.(2012){Tamm}, {Tempel}, {Tenjes}, {Tihhonova}, and
  {Tuvikene}]{tamm}
A.~{Tamm}, E.~{Tempel}, P.~{Tenjes}, O.~{Tihhonova}, and T.~{Tuvikene}.
\newblock {Stellar mass map and dark matter distribution in M 31}.
\newblock \emph{\aap}, 546:\penalty0 A4, October 2012.
\newblock \doi{10.1051/0004-6361/201220065}.

\bibitem[{Sofue}(2015)]{sofue}
Y.~{Sofue}.
\newblock {Dark halos of M 31 and the Milky Way}.
\newblock \emph{\pasj}, 67:\penalty0 75, August 2015.
\newblock \doi{10.1093/pasj/psv042}.

\bibitem[{Blumenthal} et~al.(1986){Blumenthal}, {Faber}, {Flores}, and
  {Primack}]{BlumenthalFaber}
G.~R. {Blumenthal}, S.~M. {Faber}, R.~{Flores}, and J.~R. {Primack}.
\newblock {Contraction of dark matter galactic halos due to baryonic infall}.
\newblock \emph{\apj}, 301:\penalty0 27--34, February 1986.
\newblock \doi{10.1086/163867}.

\bibitem[{Ryden} and {Gunn}(1987)]{RydenGunn}
B.~S. {Ryden} and J.~E. {Gunn}.
\newblock {Galaxy formation by gravitational collapse}.
\newblock \emph{\apj}, 318:\penalty0 15--31, July 1987.
\newblock \doi{10.1086/165349}.

\bibitem[{Cirelli} and {Panci}(2009)]{CirelliIC}
M.~{Cirelli} and P.~{Panci}.
\newblock {Inverse Compton constraints on the Dark Matter e$^{}$ excesses}.
\newblock \emph{Nuclear Physics B}, 821:\penalty0 399--416, November 2009.
\newblock \doi{10.1016/j.nuclphysb.2009.06.034}.

\bibitem[{Courteau} et~al.(2011){Courteau}, {Widrow}, {McDonald},
  {Guhathakurta}, {Gilbert}, {Zhu}, {Beaton}, and {Majewski}]{courteau}
S.~{Courteau}, L.~M. {Widrow}, M.~{McDonald}, P.~{Guhathakurta}, K.~M.
  {Gilbert}, Y.~{Zhu}, R.~L. {Beaton}, and S.~R. {Majewski}.
\newblock {The Luminosity Profile and Structural Parameters of the Andromeda
  Galaxy}.
\newblock \emph{\apj}, 739:\penalty0 20, September 2011.
\newblock \doi{10.1088/0004-637X/739/1/20}.

\bibitem[{Blumenthal} and {Gould}(1970)]{Blumenthal}
G.~R. {Blumenthal} and R.~J. {Gould}.
\newblock {Bremsstrahlung, Synchrotron Radiation, and Compton Scattering of
  High-Energy Electrons Traversing Dilute Gases}.
\newblock \emph{Reviews of Modern Physics}, 42:\penalty0 237--271, 1970.
\newblock \doi{10.1103/RevModPhys.42.237}.

\bibitem[{Rybicki} and {Lightman}(1979)]{Rybicki}
G.~B. {Rybicki} and A.~P. {Lightman}.
\newblock \emph{{Radiative processes in astrophysics}}.
\newblock 1979.

\bibitem[{Abazajian} and {Keeley}(2016)]{abazajian}
K.~N. {Abazajian} and R.~E. {Keeley}.
\newblock {Bright gamma-ray Galactic Center excess and dark dwarfs: Strong
  tension for dark matter annihilation despite Milky Way halo profile and
  diffuse emission uncertainties}.
\newblock \emph{\prd}, 93\penalty0 (8):\penalty0 083514, April 2016.
\newblock \doi{10.1103/PhysRevD.93.083514}.

\bibitem[{Ackermann} et~al.(2015)]{Ackermann2015}
M.~{Ackermann} et~al.
\newblock {Searching for Dark Matter Annihilation from Milky Way Dwarf
  Spheroidal Galaxies with Six Years of Fermi Large Area Telescope Data}.
\newblock \emph{Physical Review Letters}, 115\penalty0 (23):\penalty0 231301,
  December 2015.
\newblock \doi{10.1103/PhysRevLett.115.231301}.

\bibitem[{Steigman} et~al.(2012){Steigman}, {Dasgupta}, and {Beacom}]{steigman}
G.~{Steigman}, B.~{Dasgupta}, and J.~F. {Beacom}.
\newblock {Precise relic WIMP abundance and its impact on searches for dark
  matter annihilation}.
\newblock \emph{\prd}, 86\penalty0 (2):\penalty0 023506, July 2012.
\newblock \doi{10.1103/PhysRevD.86.023506}.

\bibitem[{Berkhuijsen} et~al.(2003){Berkhuijsen}, {Beck}, and
  {Hoernes}]{RadioData}
E.~M. {Berkhuijsen}, R.~{Beck}, and P.~{Hoernes}.
\newblock {The polarized disk in M 31 at lambda 6 cm}.
\newblock \emph{\aap}, 398:\penalty0 937--948, February 2003.
\newblock \doi{10.1051/0004-6361:20021710}.

\bibitem[{Gie{\ss}{\"u}bel} et~al.(2013){Gie{\ss}{\"u}bel}, {Heald}, {Beck},
  and {Arshakian}]{M31_350}
R.~{Gie{\ss}{\"u}bel}, G.~{Heald}, R.~{Beck}, and T.~G. {Arshakian}.
\newblock {Polarized synchrotron radiation from the Andromeda galaxy M 31 and
  background sources at 350 MHz}.
\newblock \emph{\aap}, 559:\penalty0 A27, November 2013.
\newblock \doi{10.1051/0004-6361/201321765}.

\bibitem[{Beck} et~al.(1980){Beck}, {Berkhuijsen}, and
  {Wielebinski}]{BeckBerkWiel}
R.~{Beck}, E.~M. {Berkhuijsen}, and R.~{Wielebinski}.
\newblock {Distribution of polarised radio emission in M31}.
\newblock \emph{\nat}, 283:\penalty0 272--275, January 1980.
\newblock \doi{10.1038/283272a0}.

\bibitem[{Beck} et~al.(1998){Beck}, {Berkhuijsen}, and
  {Hoernes}]{BeckBerkHoernes}
R.~{Beck}, E.~M. {Berkhuijsen}, and P.~{Hoernes}.
\newblock {A deep lambda 20 CM radio continuum survey of M 31}.
\newblock \emph{\aaps}, 129:\penalty0 329--336, April 1998.
\newblock \doi{10.1051/aas:1998187}.

\bibitem[{Beck} and {Graeve}(1982)]{BeckGrave}
R.~{Beck} and R.~{Graeve}.
\newblock {The distribution of thermal and nonthermal radio continuum emission
  of M31}.
\newblock \emph{\aap}, 105:\penalty0 192--199, January 1982.

\bibitem[{Walterbos} and {Graeve}(1985)]{WalterbosGrave}
R.~A.~M. {Walterbos} and R.~{Graeve}.
\newblock {Radio continuum emission from the nuclear region of M31 Evidence for
  a nuclear radio spiral}.
\newblock \emph{\aap}, 150:\penalty0 L1--L4, September 1985.

\bibitem[{Berkhuijsen} et~al.(1983){Berkhuijsen}, {Wielebinski}, and
  {Beck}]{BerkWielBeck}
E.~M. {Berkhuijsen}, R.~{Wielebinski}, and R.~{Beck}.
\newblock {A radio continuum survey of M31 at 4850 MHz. I - Observations - List
  of sources}.
\newblock \emph{\aap}, 117:\penalty0 141--144, January 1983.

\bibitem[{Li} et~al.(2016){Li}, {Huang}, {Yuan}, and {Xu}]{LiM31}
Z.~{Li}, X.~{Huang}, Q.~{Yuan}, and Y.~{Xu}.
\newblock {Constraints on the dark matter annihilation from Fermi-LAT
  observation of M31}.
\newblock \emph{\jcap}, 12:\penalty0 028, December 2016.
\newblock \doi{10.1088/1475-7516/2016/12/028}.

\end{thebibliography}
\end{document}